\documentclass[11pt]{article}

\def\mytitle#1{\setcounter{equation}{0}
\setcounter{footnote}{0}
\begin{flushleft}\Large\textbf{#1}\end{flushleft}
\vspace{0.25cm}}
\def\myname#1{\leftline{{\large #1}}\vspace{-0.13cm}}
\def\myplace#1#2{\small\begin{flushleft}\textit{#1}\\
\texttt{#2}\end{flushleft}}

\def\myclassification#1{\small\noindent
PACS No :
       #1\vspace{0.5cm}}
\usepackage{graphicx}
\begin{document}

\mytitle{Expanding Universe : Thermodynamical Aspects From Different Models}

\vskip0.2cm
\myname{Sridip Pal~$*$\footnote{sridippaliiser@gmail.com}}\vskip0.2cm
\myname{Ritabrata Biswas~$\dag$\footnote{biswas.ritabrata@gmail.com,
ritabrata@physics.iisc.ernet.in}} \vskip0.2cm 
\myplace{$*$ Department of Physical Sciences, Indian Institute of Science
Education and Research, Mohanpur - 741252,
Nadia, West Bengal, India}{}
\myplace{$\dag$ Department of Physics, Indian Institute of Science,
Bangalore-560012, India.}{}
\begin{abstract}
The pivotal point of the paper is to discuss the behavior of temperature,
pressure, energy density as a function of volume along with determination of
caloric EoS from following two model: $w(z)=w_{0}+w_{1}\ln(1+z)$ \& $
w(z)=-1+\frac{(1+z)}{3}\frac{A_{1}+2A_{2}(1+z)}{A_{0}+2A_{1}(1+z)+A_{2}(1+z)^{2}
}$. The time scale of instability for this two models is discussed. In the paper
we then generalize our result and arrive at general expression for energy
density irrespective of the model. The thermodynamical stability for both of the
model and the general case is discussed from this viewpoint. We also arrive at a
condition on the limiting behavior of thermodynamic parameter to validate the
third law of thermodynamics and interpret the general mathematical expression of
integration constant $U_{0}$ (what we get while integrating energy conservation
equation) physically relating it to number of micro states. The constraint on
the allowed values of the parameters of the models is discussed which ascertains
stability of universe. The validity of thermodynamical laws within apparent and
event horizon is discussed.
\end{abstract}
\myclassification{05.70.-a (Thermodynamics), 95.30.Tg (Thermodynamic processes,
conduction, convection, equations of state)}
\newpage
\tableofcontents
\section{Introduction}
To explain the cosmic acceleration predicted from the Ia type supernova
observations (Perlmutter, S. et al. 1999; Riess, A. G. et al. 1998)
 one popular wayout is to modify the stress energy tensor part, i.e., the right
hand side of the Einstein's field equation. Existence of 
some unknown matter termed as Dark Energy(DE hereafter) is been assumed (Riess,
A. G. et al 2004; Perlmutter, S. et al 1998;  
Garnavich, P. M. et al 1998; Bachall, N. A.  et. al. 1999; Copeland, E. J. et.
al. 2006)  which violates the strong energy condition.
 The simplest candidate of DE is a tiny
positive cosmological constant($\Lambda$) which obeys the equation of state (EoS
hereafter), $w=-1$. But
due to low energy scale than the normal scale for constant
$\Lambda$, the dynamical $\Lambda$ was introduced (Caldwell, R. R. et. al.
1998).
Again at very early stage of universe the energy scale for varying
$\Lambda$ is not sufficient. So to avoid this problem, known as
cosmic coincidence (Steinhardt, P. J. et. al.  1999), a new field, called
tracker field
(Zlatev, I. et. al. 1999) was prescribed. In similar way there are many models
(Sahni, V. and Starobinsky, A. A. 2003)  in Einstein gravity to best fit the
data. Yet its
require some modifications. From this point of view some
alternative models are evolved. Most of the DE models
involve one or more scalar fields with various actions and with or
without a scalar field potential (Maor, I. et. al. 1998). Now, as
the observational data permits us to have a rather time varying
EoS, there are a bunch of models characterized by
different scalar fields such as a slowly rolling scalar field
(Quintessence) ($-1<\omega<-1/3$, $\omega(=p/\rho)$, being the EoS parameter)
(Caldwell, R. R. et. al. 1998), k-essence
(Armendariz - Picon et. al. 2000), tachyon (Sen, A. 2002), phantom ($\omega<-1$)
(Caldwell, R. R. 2002), ghost condensate
 (Arkani-Hamed, N. et. al. 2004; Piazza, F and Tsujikawa, S. 2004), quintom (
Feng, B. et. al. 2005), Chaplygin gas models
 (Kamenshchik, A. Y. et. al. 2001) etc. Some recent reviews on DE models are
described in the ref. (Copeland, E.J. et. al. 2006; Li, M. et. al.2011).

While explaining evolution of the universe, various DE models have been
proposed, all of which must be constrained by astronomical
observations. In all the models, the EoS parameter $\omega$ plays a
key role and can reveal the nature of DE which
accelerates the universe. Different equations of state lead to
different dynamical changes and may influence the evolution of the
universe. The EoS parameter $\omega$ and its time
derivative with respect to Hubble time are currently constrained
by the distance measurements of the type Ia supernova and the
current observational data constrain the range of equation of
state as $-1.38<\omega<-0.82$ (Melchiorri, A. et. al. 2003). Recently, the
combination
of WMAP and Supernova Legacy Survey data shows a significant
constraint on the EoS
$\omega=-0.97^{+0.07}_{-0.09}$ for the DE, in a flat universe
(Seljak, U. et. al. 2006). Recently some parametrization for the variation of
EoS
parameters $\omega(z)$ have been proposed describing the DE component. 

1. $\omega(z)=\omega_{0}+\omega_{1}z$ (Cooray, A. R. and Huterer, D. 1999). Here
$w_{0}=-1/3$ and
$w_{1}=-0.9$ with $z<1$. This grows increasingly unsuitable for $z>1$. So the
following model has been proposed.
 We will call this parametrization as $`linear~ parametrization'$.

2. $\omega(z)=\omega_{0}+\omega_{1}\frac{z}{1+z}$. This ansatz was first
discussed by Chevallier and Polarski (Chevallier, M. and Polarski, D. 2001) and
later studied
more elaborately by Linder (Linder, E. V. 2003). The best fit values for
this model while fitting with the SNIa gold dataset are
$\omega_{0} = -1.58$ and $\omega_{1} = 3.29$. We will call this parametrization
as $`CPL(Chevallier-Polarski-Linder)~ parametrization'$.

3. $\omega(z)=\omega_{0}+\omega_{1}\frac{z}{(1+z)^{2}}$ ( Jassal, H. K. et. al.
2005). A fairly rapid evolution of this EoS allowed
so that $\omega(z)\ge -1/2$ at $z>0.5$ is consistent with the supernovae
observation.
 We will call this parametrization as $`JBP(Jassal-Bagla-Padmanabhan)~
parametrization'$.

4. $ \omega(z)=-1+\frac{(1+z)}{3}
\frac{A_{1}+2A_{2}(1+z)}{A_{0}+2A_{1}(1+z)+A_{2}(1+z)^{2}}$
(Alam, U. et. al. 2004a, 2004b). This ansatz is exactly the cosmological
constant
$\omega = -1$ for $A_{1} = A_{2} = 0$ and DE models with $w =-2/3$ for
$A_{0} = A_{2} = 0$ and $\omega = -1/3$ for $A_{0} = A_{1} = 0$. It has
also been found to give excellent results for DE models in which
the equation of state varies with time including quintessence,
Chaplygin gas, etc. The best fit values of $A_{1}$ and $A_{2}$ are $A_{1}=-4.16$
and $A_{2}=1.67$ for the SN1a Gold dataset.
 We will call this parametrization as $`ASSS(Alam-Sahni-Saini-Starobinski)~
parametrization'$.

5. $\omega(z)=\omega_{0}+\omega_{1}\ln(1+z)$ (Efstathiou, G. 1999). This
evolution form of EoS is valid for $z<4$.
$\omega_{1}$ is a small number which can be determined by the observations. The
minimum value of $\omega_{1}$ is
approximately $-0.14$ and $\omega_{0}\ge -1$. We will call it
$`Log~parametrization'$.\\

Thermodynamics of DE universe filled up with the fluids with linear and JBP
parametrization had been studied before
 (Mazumder, N. et. al. 2010). Thermodynamics with CPL was studied by Xing, L.
et. al. (2011). In this paper we are 
going to study DE universe filled with fluid Log parametrization and ASSS
parametrization respectively which have
 not been studied before so extensively.

As we went through the literature we came to know that apart from the
cosmological constant ($\omega=-1$), the 
sinehyperbolic scalar field potential (Sahni, V. and Starobinsky, A.A 2000;
Urena-Lopez, L.A. and Matos, T. 2000; 
Sahni, V. et al. 2003) and the topological defect models, there is no DE model
with constant $\omega$ consistent with observation 
(Alam, U. et. al. 2004a). Almost every model like Quintessence, Chaplygin gas
(Setare, M.R. and Bouhmadi-Lopez, M.R. 2007, Setare, M.R. 2007b)
 etc. depicts significant evolution in $\omega(z)$ over sufficient time scale.

The way towards a meaningful reconstruction of $\omega(z)$ depends on inventing
an efficient fitting function
 for either $d_{L}(z)$ or $H(z)$, where $d_{L}(z)$ is luminosity distance and
$H(z)$ is Hubble parameter. 
The parameters of this fitting function are determined by matching to Supernova
observations. Now, one can
 manage to reconstruct the functional form of $\omega(z)$ by taking an ansatz of
the luminosity distance or
 $H(z)$ and then by comparing it to supernova observations.\\

This model is based on the following ansatz:
\begin{equation}
H(x)=H_{0}\left(\Omega_{m}x^{3}+A_{0}+A_{1}x+A_{2}x^{2}\right)^{\frac{1}{2}}.
\end{equation}
Which can be equivalently written as following:
\begin{equation}
\rho_{DE}=\rho_{0c}\left(A_{0}+A_{1}x+A_{2}x^{2}+A_{3}x^{3}\right)
\end{equation}
where $x=\left(1+z\right)$, $\rho_{0c}$ is the critical density at present time
and $\Omega_{m}$ dimensionless density of dark energy.\\

If we carefully note we can see the ansatz in terms of energy density is
actually a truncated taylor expansion where it is exploited that any well
bahaved function can be well approximated by Taylor expansion within a range. So
the model is valid for $z \leq $ few (Alam U et. al. 2004a).

The model yields excellent results among DE models in which the EoS varies with
time including quintessence, Chaplygin gas etc. (Sahni, V. et al. 2003; Alam, U.
et al. 2003). Alam, U et al. 2004a (In figure 1 of their paper) have shown the
accuracy of the ansatz when it is applied to several other DE models like
tracker quintessence, the Chaplygin gas and super-gravity (SUGRA) models. They
also plotted the deviation of $\log\left(d_{L}H_{0}\right)$ (which is the
measured quantity for SNe) obtained via the ansatz from the actual model values.
Clearly the ansatz comes out to be of excellent agreement well over a
significant red-shift range for $\Omega_{0m}=0.3$. The ansatz agrees with these
models of dark energy with less than $0.5\%$ errors in the redshift range where
we do have SNe data available (Alam, U. et. al. 2004a)

The analysis of type Ia supernova data involving the priors with most frequently
used condition as $\omega$ = constant and $\omega \geq−1$ leads to confinement
of DE to within a narrow class of models. Moreover, when we impose such priors
on the cosmic equation of state it can result to a complete misrepresentations
of reality as shown in (Maor, et al. 2002).

Recently, evolution of EoS parameter with red shifts been investigated (Vazquez,
J. A. et. al. 2012) by performing a Bayesian analysis of current cosmological
observations. Vazquez, J. A. et. al. have shown if they calculate the Bayes'
factor then most of the data catalogues supports the fact that CPL, JBP are
significantly disfavored with respect to simple $\Lambda CDM$ model
($\omega=-1$). Now, linear parametrization is not to be considered in such a
case as the parametrization blows with high $z$. Calculation of Bayes' factor
for ASSS and $\Lambda CDM$ using codes like CAMB etc are bit cumbersome. But
from intuition we can see $-1$ term is already there in ASSS parametrization.
Extra additional $z$ dependent term may make it different enough from the
$\Lambda CDM$ model. Even theoretically, ASSS is efficient enough to explain
different phases of universe. These observational supports substantiate our
interest of working with ASSS model.

Linear, CPL, JBP or ASSS - all the parametrizations can be treated as kind of
combinations of polynomials with $(1+z)$ as polynomial parameter. These modes
are efficient for early epochs whereas they are not that effective for low
redshifts. While parametrizing, the fact that at low redshifts the
magnitude-redshift relation is degenrate for the models having same deceleration
parameter, should be taken into account. Besides keeping consistency of SN
constraints on dimensionless DE density with those derived from CAMB
measurements is another important factor. These constrains along with many
others had given birth of Log parametrization. This is somehow not following the
common pathway of considering DE EoS. These properties did motivate us to take
Log parametrization for the study of thermodynamics of universe.

We will study the universe from thermodynamical aspect for both type of fluids
separately in Sec. \ref{section2}. In Sec. \ref{section3} we will seek the
thermal EoS for both type of fluid separately along with study on
thermodynamical stability. The Internal Energy, thermodynamic pressure, entropy,
temperaure as a function of volume have been determined in this section. The
Sec. \ref{section4} contains a general derivation of thermal EoS for any fluid
obeying $p=\omega(z) \rho$ and a general discussion on stability criterion. We
have tried to give a physical interpretation of the generality of thermal EoS
and the expression for energy density as a function of temperature and volume in
this section. Study of instable cases, onset of instability have been discussed
in Sec. \ref{sectionsuppliment}. The Sec. \ref{section5} proves the validity of
laws of thermodynamics on the apparent horizon and invalidity of the same on the
event horizon which agrees with (Mazumder, N. et. al. 2010; Xing, L. et. al.
2011). The paper ends with a brief concluding remark in Sec. \ref{section6}.
\section{Study of Universe Treating it as an Thermodynamical
System}\label{section2}
Let us consider an Universe filled with a perfect fluid having volume $V$ and
$\rho,~p,~T$ and $S$ are
respectively the energy density, thermodynamical pressure,
temperature and entropy of the system. From the first law of
thermodynamics (Myung, Y. S. 2011) we have
\begin{equation}\label{1}
TdS=d(\rho V)+pdV=d[(\rho + p)V]-Vdp
\end{equation}
We have the following integrability condition (Myung, Y. S. 2011), i.e., $$
\frac{\partial^{2}S}{\partial T
\partial V} = \frac{\partial^{2}S}{\partial V \partial T}~$$ which yields
(Gong, Y. et. al. 2007)
\begin{equation}\label{2}
\frac{dp}{\rho + p}=\frac{dT}{T}.
\end{equation}
So combining  equations (\ref{1}) and (\ref{2}) and integrating we obtain the
following:
\begin{equation}\label{3}
S=\frac{(\rho + p)V}{T},
\end{equation}
where we have dropped an additive constant devoid of physical significance.\\

We assume our universe to be homogeneous and isotropic FRW
space-time with following line element
\begin{equation}\label{4}
ds^{2}=-dt^{2}+a^{2}(t)\left[\frac{dr^{2}}{1-kr^{2}}+~{r}^{2}\left(d\theta^{2}
+\sin^{2}\theta d\phi^{2}\right) \right],
\end{equation}
where $k$, the curvature scalar
having values $0,~\pm 1$ for flat, closed and open universe respectively.
The Friedmann equations and the energy conservation equation are
\begin{equation}\label{5}
H^{2}=\frac{8\pi G \rho}3 - \frac{k}{a^{2}}~,
\end{equation}
\begin{equation}\label{6}
\dot{H}=-4\pi G (\rho + p) + \frac{k}{a^{2}}
\end{equation}
and
\begin{equation}\label{7}
\dot{ \rho }+3H(\rho + p) = 0~,
\end{equation}
where the Hubble parameter is given by $H=\frac{\dot{a}}a$.
\subsection{{\bf Thermodynamics of Fluid with
$\omega(z)=\omega_{0}+\omega_{1}\ln \left(1+z\right)$}}
Using this $\omega(z)$, integrating the energy conservation equation (\ref{7})
we would get the expression of energy density as a function of redshift as
follows :
\begin{equation}\label{8}
\rho=\rho_{0}(1+z)^{3\left\{1+\omega_{0}+\frac{\omega_{1}}{2}ln(1+z)\right\}},
\end{equation}
where $\rho_{0}$ is integration constant denoting the present time ($z=0$)
density of our universe. \\

\begin{enumerate}

\item From the graph of energy density vs redshift, we observe that $\rho$
initially has a very high value. With expansion of the universe, $\omega$
decreases, hence $\rho$ falls to $\rho_{0}$ as $z$ becomes $0$ from $2$.

\item The graph of Pressure vs Redshift reveals that the pressure remains
positive for allowed values of $\omega_{0}$ and $\omega_{1}$ in the range $ z
\in (0,2)$. It initially starts with high value and then decreases down to
present value as $z$ goes  from $2$ to $0$.

\end{enumerate}

The integration of the integrability condition (\ref{2}) gives expression for
temperature as
\begin{equation}\label{9}
T=T_{0}(1+\omega)(1+z)^{3\left\{\omega_{0}+\frac{\omega_{1}}{2}ln(1+z)\right\}},
\end{equation}
where $T_{0}$ is the integration constant(Note $T_{0}(1+\omega_{0})$ is the
present time temperature).\\
\begin{itemize}
\item The graph between Temperature and Redshift reveals that Temperature is
also dropping as we traverse from past to present, i.e, as $z$ varies from $2$
to $0$.
\end{itemize}
Using the last two expressions and plugging in the expression for entropy in the
last section, we get the following expression for $S$ :
\begin{equation}\label{10}
S=\frac{\rho_{0}}{T_{0}}.
\end{equation}
The heat capacity and square of the sound velocity are given by:
\begin{equation}\label{11}
c_{V}(z)=V\frac{\partial{\rho}}{\partial{T}}=\frac{3S\left(1+\omega\right)}{
\omega_{1}+3\omega\left(1+\omega\right)}.
\end{equation}\\
Note when $\omega < -1$ we have $c_{V} < 0$ i.e universe is in a unstable phase.
Hence the stability condition demands:
$\omega > -1 $. 

The expression for the sound's speed :
\begin{equation}\label{12}
\mathit{v}_{s}^{2}(z)=\frac{\partial{p}}{\partial{\rho}}=\omega+\frac{\omega_{1}
}{3(1+\omega)} .
\end{equation}
Now demanding $\mathit{v}_{s}^{2}<1$ we get the following condition:
\begin{equation}\label{13}
\left\{\omega_{0}+\omega_{1}ln(1+z)\right\}^{2} \leq \frac{3-\omega_{1}}{3},
\end{equation}
along with the constraint $\omega_{1}\leq 3$. $z$ has a minimum value of $0$,
hence we can write:
\begin{equation}\label{14}
3\omega_{0}^{2}+\omega_{1} \leq 3.
\end{equation}
\begin{figure}
~~~~~~~~~~~~~~~Fig 1~~~~~~~~~~~~~~~~~~~~~~~~~~~~~~~~~~~~Fig
2~~~~~~~~~~~~~~~~~~~~~~~~~~~~~~~~~~~~Fig 3\\
\includegraphics[height=1.8in,
width=1.8in]{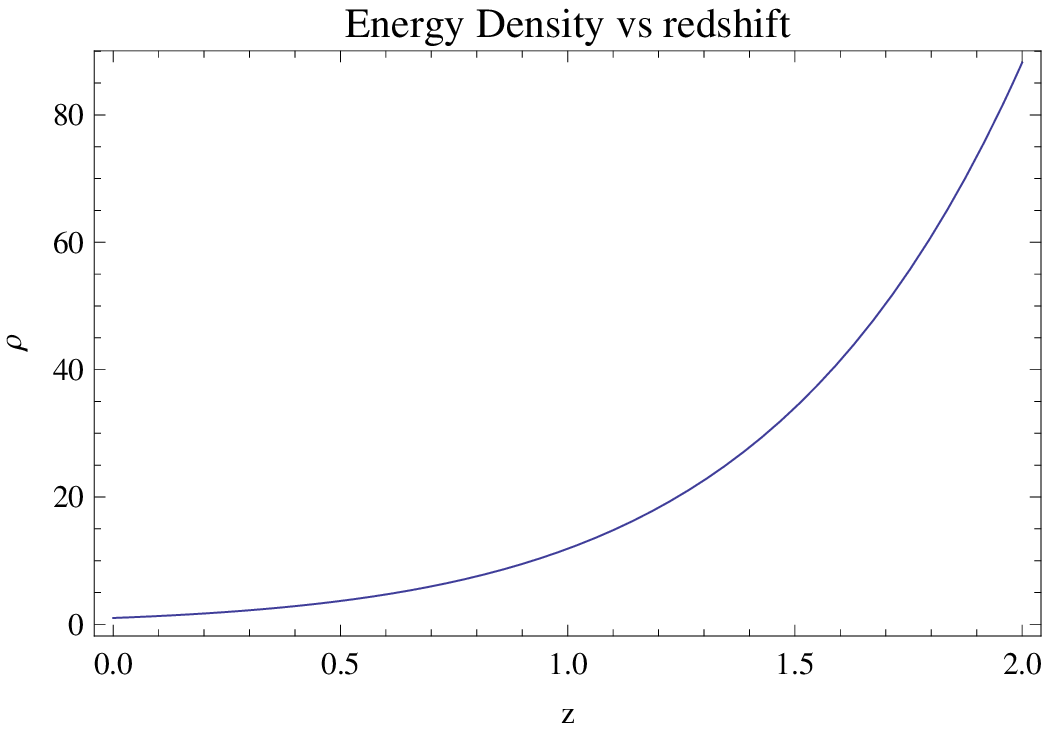}~~~~~~~\includegraphics[height=1.8in,
width=1.8in]{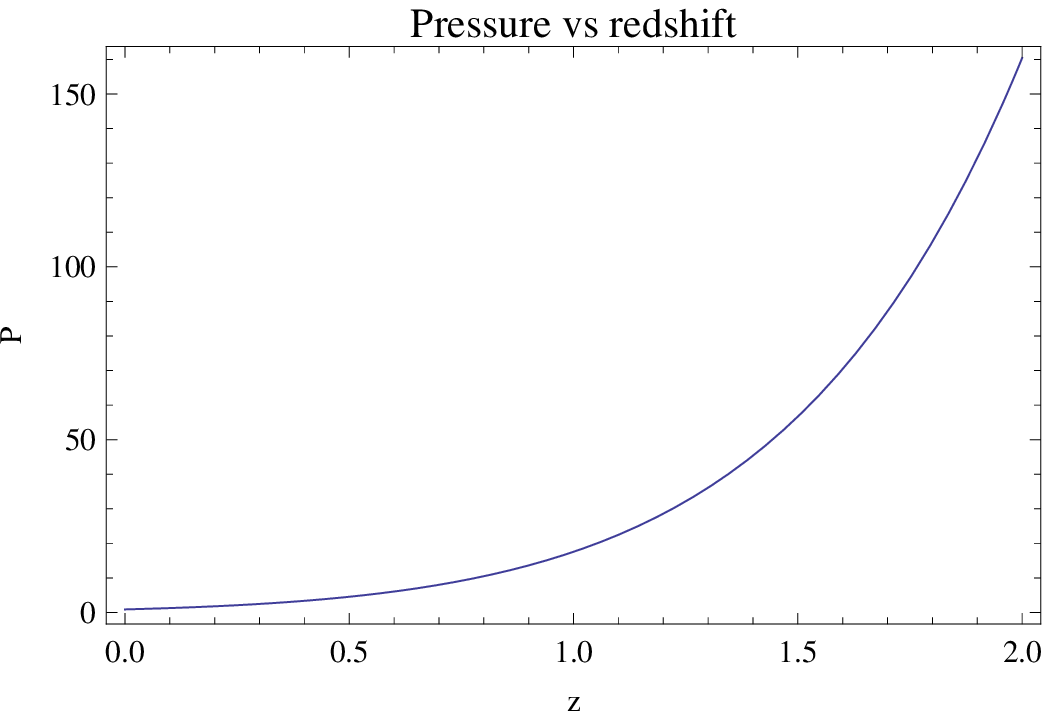}
~~~~~~~\includegraphics[height=1.8in, width=1.8in]{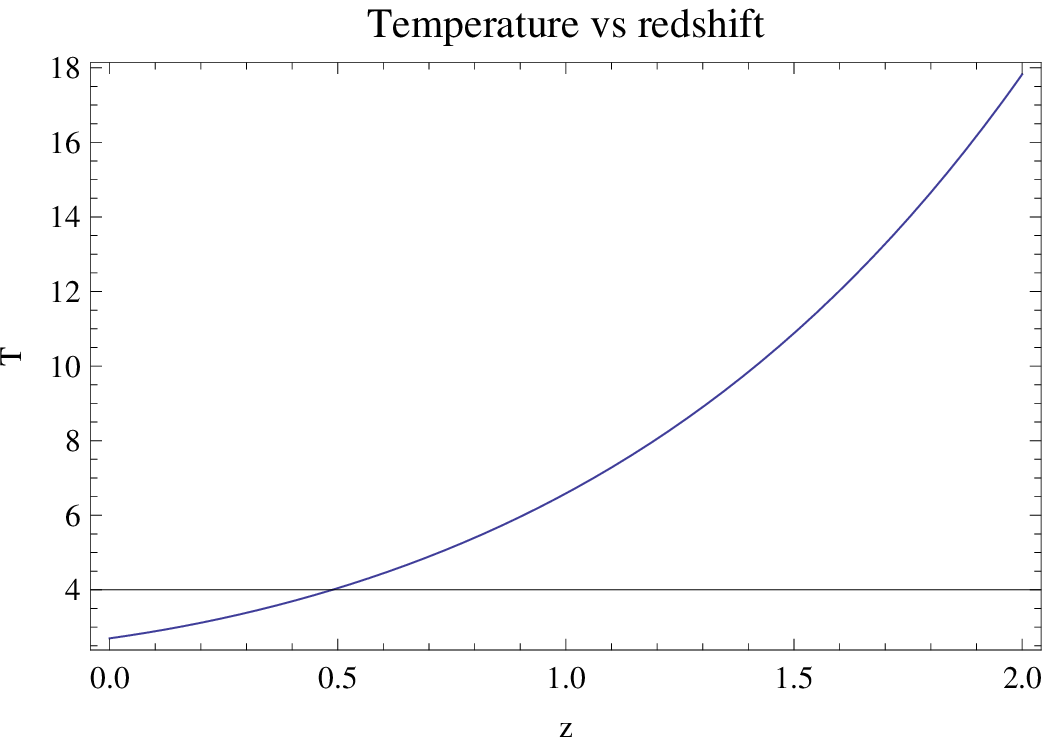}
\end{figure}
\subsection{{\bf Thermodynamics of Fluid with
$\omega(z)=-1+\frac{1+z}{3}\frac{A_{1}+2A_{2}(1+z)}{A_{0}+2A_{1}(1+z)+A_{2}
(1+z)^{2}}$}}
Using this $\omega(z)$, integrating the energy conservation equation we would
get the following:
\begin{equation}\label{15}
\rho=\rho_{0}M\exp{\left\{-\frac{A_{1}}{\sqrt{A_{0}A_{2}-A_{1}^{2}}}\arctan{
\left(\frac{A_{0}+A_{2}(1+z)}{\sqrt{A_{0}A_{2}-A_{1}^{2}}}\right)}\right\}},
\end{equation}
where $\rho_{0}$ is integration constant which we can get putting $z=0$, and
$M=A_{0}+2A_{1}(1+z)+A_{2}(1+z)^{2}$.

On the other hand integrating the integrability condition, we get the expression
for temperature:
\begin{equation}\label{16}
T=T_{0}(1+\omega)(1+z)^{-3}M\exp{\left\{-\frac{A_{1}}{\sqrt{A_{0}A_{2}-A_{1}^{2}
}}\arctan{\left(\frac{A_{0}+A_{2}(1+z)}{\sqrt{A_{0}A_{2}-A_{1}^{2}}}\right)}
\right\}},
\end{equation}
where $T_{0}$ is the integration constant.

Using the last two expressions and plugging in the expression for entropy in the
last section, we get the following expression for $S$:
\begin{equation}\label{17}
S=\frac{\rho_{0}}{T_{0}}.
\end{equation}

The heat capacity and square of the sound velocity are given by:
\begin{equation}\label{18}
c_{V}(z)=V\frac{\partial{\rho}}{\partial{T}}=
S\left\{\frac{1}{1+\omega}+\frac{\frac{3}{(1+\omega)(1+z)}-\frac{1}{(1+\omega)^{
2}}\frac{\partial\omega}{\partial z}}{\frac{1}{1+\omega}\frac{\partial
\omega}{\partial{z}}+\frac{3\omega}{1+z}}\right\}.
\end{equation}
Upon simplification we get:
\begin{equation}\label{18.1}
c_{V}=\frac{3S\left(2+\omega\right)}{3\left(1+\omega\right)^{2}
+\left(1+z\right)\frac{\partial \omega}{\partial z}}~~~~and
\end{equation}\\
\begin{equation}\label{19}
\mathit{v}_{s}^{2}(z)=\frac{\partial{p}}{\partial{\rho}}\\=\omega+\frac{\partial
{w}}{\partial{z}}\left\{\frac{A_{1}+A_{2}(1+z)}{A_{0}+2A_{1}(1+z)+A_{2}(1+z)^{2}
}\right\}^{-1}
\end{equation}
\section{Derivation of Thermal EoS \& Study of Stability}\label{section3}
In this approach it is very convinient to consider the fluid obeying adiabatic
EoS $p=\omega (z) \rho$ with constant particle number $N$ as a thermodynamical
system. Without any loss of generality we can assume the internal energy ($U$)
and the pressure ($p$) as functions of entropy ($S$) and volume ($V$). So we can
structurize our density and pressure and also the concerend differential
equation as (Landau, L. D. and Lifschitz, E. M. 1984)
\begin{eqnarray}\label{20}
\left.\begin{array}{c}
\rho=\frac{U}{V},\\\\
p=-\left(\frac{\partial U}{\partial V}\right)_{S}~~~~and\\\\
\frac{dU}{dV}+\omega (z)\frac{U}{V}=0.
\end{array}\right\}
\end{eqnarray} 
\subsection{{\bf Fluid with $\omega(z)=\omega_{0}+\omega_{1}\ln
\left(1+z\right)$}}
In (\ref{20}), use of $\omega(z)=\omega_{0}+\omega_{1}ln(1+z)$, yields the
following:
\begin{equation}\label{22}
U=U_{0}V^{(\frac{\omega_{1}}{6}lnV)}\exp{(-\omega_{0}V)},
\end{equation}
where $U_{0}$ is integration constant (It can be function of $S$ and
$\omega_{0}$ should be greater than $0$ for $U$ not to diverge). Hence the
energy density becomes
\begin{equation}\label{23}
\rho= U_{0}V^{(\frac{\omega_{1}}{6}lnV-1)}\exp{(-\omega_{0}V)},
\end{equation}
while the expression for pressure is following:
\begin{equation}\label{24}
p=U_{0}\left\{\omega_{0}-\frac{\omega_{1}}{3}lnV\right\}V^{(\frac{\omega_{1}}{6}
lnV-1)}\exp{(-\omega_{0}V)}.
\end{equation}
The criterion for stability of the fluid during expansion:
\begin{enumerate}
\item $\left(\frac{\partial p}{\partial V}\right)_{S}< 0$.
\item The thermal capacity at constant volume should be greater than zero i.e
$c_{V}> 0$.
\end{enumerate}
The first condition leads to the following constraint:
\begin{equation}\label{24.1}
\left(\frac{\omega_{1}}{3}lnV-1-\omega_{0}V\right)\left(\omega_{0}-\frac{\omega_
{1}}{3}lnV\right)<\frac{\omega_{1}}{3}.
\end{equation}\\

Now to get to the thermal equation of state we start with the expression for
temperature:
\begin{equation}\label{25}
T=\left(\frac{\partial U}{\partial S}\right)_{V}.
\end{equation}
Using the expression for internal energy we obtain:
\begin{equation}\label{26}
T=\frac{U}{U_{0}}\frac{dU_{0}}{dS}=\frac{dU_{0}}{dS}V^{(\frac{\omega_{1}}{6}lnV)
}\exp{(-w_{0}V)}.
\end{equation}

Now we will look into $U_{0}$ by considering the change of variable from (S,V)
to (P,T). The Jacobian of the transformation is (Landau, L. D. and Lifschitz, E.
M. 1984):
\begin{equation}\label{27}
J=\left(\frac{\partial p}{\partial S}\right)_{V}\left(\frac{\partial T}{\partial
V}\right)_{S}-\left(\frac{\partial p}{\partial V}\right)_{S}\left(\frac{\partial
T}{\partial S}\right)_{V}.
\end{equation}
Note,we can also write it out as the following:
\begin{equation}\label{27.1}
J=-\left(\frac{\partial T}{\partial S}\right)_{V} \left(\frac{\partial
p}{\partial S}\right)_{T}.
\end{equation}
Equating the last two expressions and using the expression for $U$ and $T$ we
get the following:
\begin{equation}\label{28}
\left(\omega_{0}-\frac{\omega_{1}}{3}\ln
V\right)\left(-\omega_{0}+\frac{\omega_{1}}{3V}\ln
V\right)\left\{U_{0}\frac{d^{2}U_{0}}{dS^{2}}-\left(\frac{dU_{0}}{dS}\right)^{2}
\right\}=0
\end{equation}
\begin{equation}\label{29}
\Rightarrow U_{0}\frac{d^{2}U_{0}}{dS^{2}}-\left(\frac{dU_{0}}{dS}\right)^{2}=0
\end{equation}
\begin{equation}\label{30}
\Rightarrow U_{0}=\exp{(\alpha S)},
\end{equation}
where $\alpha$ is integration constant upon integrating the differential
equation for $U_{0}$. Hence we find
\begin{equation}\label{31}
\rho=\frac{T}{\alpha V}~~~\&
\end{equation}
\begin{equation}\label{32}
p=\omega \rho =\frac{T}{\alpha
V}\left(\omega_{0}-\frac{1}{3}\omega_{1}lnV\right)
\end{equation}\\
The existence of critical point requires the following condition to be true:
\begin{eqnarray}\label{32.1}
\left.\begin{array}{c}
\left(\frac{\partial p}{\partial V}\right)_{T}=0\\\\
\left(\frac{\partial^{2}p}{\partial V^{2}}\right)_{T}=0\\\\
\left(\frac{\partial^{3}p}{\partial V^{3}}\right)_{T} < 0
\end{array}\right\}
\end{eqnarray}\\
But in this model, the point where we have $\left(\frac{\partial p}{\partial
V}\right)_{T}=0$ has the property that second of above criterion does not hold
as
\begin{equation}\label{32.2}
\left(\frac{\partial^{2}p}{\partial
V^{2}}\right)_{T}=\frac{T}{\alpha}\frac{\omega^{\prime\prime}(V)}{V}=\frac{
\omega_{1}}{3V^{2}}\frac{T}{\alpha V}\neq 0~~,
\end{equation}
where prime($\prime$) denotes derivative with respect to volume. Hence, in this
model universe does not go through any critical point, even if there is a
transition from unstable to stable configuration or vice-versa the transition is
smooth.\\
The specific heat at constant volume comes out to be :
\begin{equation}\label{33}
c_{V}=T\left(\frac{\partial S}{\partial T}\right)_{V}=\frac{1}{\alpha}~~.
\end{equation}
Choosing $\alpha >0$ guarantees the fulfillment of second criterion for
stability, i.e., $c_{V} >0$.
while the velocity of sound is given by:
\begin{equation}\label{34}
\mathit{v}_{s}^{2}=\omega - \frac{\omega_{1}}{\omega_{1}lnV -3 -3\omega_{0}
V}~~.
\end{equation}
Here also $\mathit{v}_{s}^{2} \leq 1$ imposes constraint on the allowed values
of $\omega_{0}$ and $\omega_{1}$.

From the expression for $U_{0}$, the entropy is given by:
\begin{equation}\label{35}
S=\frac{1}{\alpha}\left[ln\left(\frac{T}{\alpha}\right)+\omega_{0}V-\frac{
\omega_{1}}{6}(lnV)^{2}\right]~~.
\end{equation}
\subsection{{\bf Fluid with
$\omega(z)=-1+\frac{1+z}{3}\frac{A_{1}+2A_{2}(1+z)}{A_{0}+2A_{1}(1+z)+A_{2}
(1+z)^{2}}$}}
The expression for $U$ in this case becomes:
\begin{equation}\label{36}
U=U_{0}\left(A_{0}V+2A_{1}V^{2/3}+A_{2}V^{1/3}\right)\exp
\left\{\frac{A_{1}~\arctan{\left(\frac{A_{0}V^{1/3}+A_{1}}{\sqrt{A_{0}A_{2}-A_{1
}^{2}}}\right)}}{\sqrt{A_{0}A_{2}-A_{1}^{2}}}\right\}
\end{equation}
where $U_{0}$ is integration constant. Therefore energy density comes out to be:
\begin{equation}\label{37}
\rho= U_{0}\left(A_{0}+2A_{1}V^{-1/3}+A_{2}V^{-2/3}\right)\exp
\left\{\frac{A_{1}~\arctan{\left(\frac{A_{0}V^{1/3}+A_{1}}{\sqrt{A_{0}A_{2}-A_{1
}^{2}}}\right)}}{\sqrt{A_{0}A_{2}-A_{1}^{2}}}\right\}
\end{equation}
while the pressure is given by
\begin{equation}\label{38}
p=\omega(V) U_{0}\left(A_{0}+2A_{1}V^{-1/3}+A_{2}V^{-2/3}\right)\exp
\left\{\frac{A_{1}~\arctan{\left(\frac{A_{0}V^{1/3}+A_{1}}{\sqrt{A_{0}A_{2}-A_{1
}^{2}}}\right)}}{\sqrt{A_{0}A_{2}-A_{1}^{2}}}\right\}
\end{equation}
where $\omega(z)=
\left(-1+\frac{1+z}{3}\frac{A_{1}+2A_{2}(1+z)}{A_{0}+2A_{1}(1+z)+A_{2}(1+z)^{2}}
\right)$ and $V=a^{3}=(1+z)^{-3}$.

Now we have two  criterions for stability of the fluid during expansion :
\begin{enumerate}
\item $\left(\frac{\partial p}{\partial V}\right)_{S}< 0$
\item The thermal capacity at constant volume should be nonnegative i.e $c_{V}>
0$
\end{enumerate}
The first condition yields:
\begin{equation}\label{38.1}
\left(A_{0}+2A_{1}V^{-\frac{1}{3}}+A_{2}V^{-\frac{2}{3}}\right)\left(\frac{
\partial \omega}{\partial
V}+\omega(V)\frac{A_{1}A_{0}V^{-\frac{2}{3}}}{3\left(A_{0}V^{\frac{1}{3}}+A_{1}
\right)^{2}+\left(A_{0}A_{2}-A_{1}^{2}\right)}\right) < \frac{2}{3}R
\end{equation}
where $R=\left(A_{1}V^{-\frac{4}{3}}+A_{2}V^{-\frac{5}{3}}\right)\omega(V)$.

Now we start with the expression for temperature to obtain thermal EoS:
\begin{equation}\label{39}
T=\left(\frac{\partial U}{\partial S}\right)_{V}
\end{equation}
Using equation(\ref{36}) we obtain:
\begin{equation}\label{40}
T=\frac{U}{U_{0}}\frac{dU_{0}}{dS}=\frac{dU_{0}}{dS}(A_{0}V+2A_{1}V^{2/3}+A_{2}
V^{1/3})\exp
\left\{\frac{A_{1}~\arctan{\left(\frac{A_{0}V^{1/3}+A_{1}}{\sqrt{A_{0}A_{2}-A_{1
}^{2}}}\right)}}{\sqrt{A_{0}A_{2}-A_{1}^{2}}}\right\}
\end{equation}

Equating (\ref{27}) and (\ref{27.1}) for this system we have,
$$
U_{0}\frac{d^{2}U_{0}}{dS^{2}}-\left(\frac{dU_{0}}{dS}\right)^{2}=0 \Rightarrow
U_{0}=\exp{(\alpha S)}$$
This result resembles with (\ref{30}). Even we can see the expression for energy
density will be similar to (\ref{31}). Expression for pressure in this case is :
\begin{equation}\label{45}
p=\omega \rho
=\left\{-1+\frac{1+z}{3}\frac{A_{1}+2A_{2}(1+z)}{A_{0}+2A_{1}(1+z)+A_{2}(1+z)^{2
}}\right\}\frac{T}{\alpha V}
\end{equation}\\
The same conditions (\ref{32.1}) applies here too which yields the following
conditions needed to be satisfied for having critical point:
\begin{equation}\label{45.1}
\omega^{\prime\prime}(V)=0\  \&\  \omega^{\prime\prime\prime}(V) < 0
\end{equation}
The thermal capacity of the system at constant volume becomes :
\begin{equation}\label{46}
c_{V}=T\left(\frac{\partial S}{\partial T}\right)_{V}=\frac{1}{\alpha}
\end{equation}
If we choose $\alpha >0$ it automatically guarantees the fulfillment of second
criterion for stability in the same way as previous section i.e $c_{V} >0$.

While the velocity of sound is given by:
\begin{equation}\label{47}
\mathit{v}_{s}^{2}=\omega +\frac{\frac{\partial \omega}{\partial
V}}{\frac{\partial (\ln
 p)}{\partial V}}
\end{equation}
Now $\mathit{v}_{s}^{2} <1 $ imposes constraint on $\omega_{0}$ and
$\omega_{1}$.

From the expression for $U_{0}$, we obtain the expression for entropy which is:
\begin{equation}\label{48}
S=\frac{1}{\alpha}\left[ln\left(\frac{T}{\alpha}\right)-\ln\left(A_{0}V+2A_{1}V^
{2/3}+A_{2}V^{1/3}\right)-\left\{\frac{A_{1}~\arctan{\left(\frac{A_{0}V^{1/3}+A_
{1}}{\sqrt{A_{0}A_{2}-A_{1}^{2}}}\right)}}{\sqrt{A_{0}A_{2}-A_{1}^{2}}}\right\}
\right]
\end{equation}
\section{A General Derivation : Independent of Model}\label{section4}
Looking deep into the fact that in last section we have observed $U_{0}$ comes
out to be $\exp{(\alpha S)}$ independent of the form of $\omega(z)$, here we are
giving a proof without assuming any specific form of $\omega(z)$ which makes the
result stronger.

We start with integrating energy conservation equation (\ref{20})
\begin{equation}\label{49}
U=U_{0}F(V)
\end{equation}
where $F(V)= \exp\left\{- {\Large \int}\frac{\omega(V)}{V} dV\right\}$.
Hence $P$ becomes:
\begin{equation}\label{50}
P=U_{0}\frac{F(V)\omega(V)}{V}
\end{equation}
Now to get to the thermal equation of state we start with the expression for
temperature (\ref{25}) and using the expression for internal energy we obtain:
\begin{equation}\label{52}
T=\frac{U}{U_{0}}\frac{dU_{0}}{dS}=\frac{dU_{0}}{dS}F(V)
\end{equation}

Equating (\ref{27}) and (\ref{27.1}) and using the expression for $U$ and $T$ we
get the following:
\begin{equation}\label{54}
\frac{F(V)F'(V)\omega
(V)}{V}\left[U_{0}\frac{d^{2}U_{0}}{dS^{2}}-\left(\frac{dU_{0}}{dS}\right)^{2}
\right]=0
\end{equation}
$$
\Rightarrow U_{0}\frac{d^{2}U_{0}}{dS^{2}}-\left(\frac{dU_{0}}{dS}\right)^{2}=0
\Rightarrow U_{0}=\exp{(\alpha S)}
$$
where $\alpha$ is integration constant upon integrating the differential
equation for $U_{0}$. Immediately, the expression for the energy density becomes
$\rho=\frac{T}{\alpha V}$ while the pressure is given by
\begin{equation}\label{57}
p=\frac{\omega T}{\alpha V} 
\end{equation}

The specific heat at constant volume comes out to be $
c_{V}=T\left(\frac{\partial S}{\partial T}\right)_{V}=\frac{1}{\alpha}
$.\\
and the entropy is given  by following expression:
\begin{equation}\label{57.1}
S=\frac{1}{\alpha}\left[ln\left(\frac{T}{\alpha}\right)-lnF(V)\right]
\end{equation}\\
So, $T$ should be $\geq \alpha F(V)$ to ascertain $S\geq 0$. 

The third law of thermodynamics demands the entropy to go to 0 as temperature
approaches zero. This demand translates into following criterion when we use
(\ref{57.1}):
\begin{equation}\label{57.11}
\lim_{T\rightarrow 0} \frac{F(V)}{T} =\frac{1}{\alpha}=c_{V}
\end{equation}\\
So we can conclude $F(V)$ must tend to 0 as universe cools down to absolute
zero. Using this information, from equation (\ref{57.11})  we can arrive at the
fact that
$F(V)$ is actually a measure of change of internal energy to effect a small
change in temperature around $T=0$ point. \\\\
{\bf Physically Interpreting The General Form of $U_{0}$ : }\\\\
Irrespective of the model we have obtained,
\begin{equation}\label{57.2}
U_{0}=\exp \left(\alpha S\right)
\end{equation}
In its differential form we can write
\begin{equation}\label{57.3}
\frac{dU_{0}}{U_{0}}=\alpha dS
\end{equation}
Now note we have $S=k_{B}ln\Omega$ where $\Omega$ is the number of micro states
corresponding to a macrostate of the system. In its differential form it looks
very much like (\ref{57.3})
\begin{equation}\label{57.4}
\frac{d\Omega}{\Omega}=\frac{1}{k_{B}}dS
\end{equation}
So comparing equation (\ref{57.3}) and (\ref{57.4}), we interpret $U_{0}$ as a
measure of number of micro states. Now $U_{0}$ is related to energy of universe
at present and it is physically very plausible to have energy going in
proportional to the number of micro states. 
\section{Instability : Time scale for Onset of
Instabilities}\label{sectionsuppliment}
He was Einstein in very known history who has made ``the blunder" while
producing a static universe which was unstable and not observationally supported!
Latter instability regarding cosmological, matter creation has been studied
(Saslaw, W. C. 1967). General idea of thermodynamics says that negative specific
heat indicates an instability(of course thermal).\\\\
{\bf Logarithmic Model}\\
Here the expression of $c_{V}$ illustrates two cases of instability:\\
$(1)$ $\omega < -1$ along with $\omega_{1}+3\omega(1+\omega) > 0$ : Now it can
be easily noted that the former criterion implies
 the later. Hence we have instability for $\omega < -1$. Taking 
$\omega_{0}=-0.995$ and $\omega_{1}=0.25$ (Efstathiou, G. 1999)
  we would get that $z=-0.019$ marks the onset of instability. As $z$ goes in
the opposite direction of time, negative $z$ indicates towards the future. So
$z=-0.019$ can be treated as the timescale for onset of instability. This may
suggest that our universe is now in a thermally stable equilibrium but is
heading towards an instability where small perturbation can lead to catastrophic
change.\\\\
$(2)$ $-1<\omega$ along with $\omega_{1}+3\omega(1+\omega) < 0$ : Now $\omega>0$
does not satisfy this condition, hence we have stability when $\omega >0 $. So
the range is narrowed down to $-1<\omega<0$ which in turn implies an interesting
scenario that universe starts off with a stable state and it retains stability
until $\omega$ hits $0$. Thereafter it will undergo a stage of thermal
instability after which again it will be in a thermally stable state when
$\omega$ drops below $-1$.\\

In short, depending on the signature of  $\omega_{1}+3\omega(1+\omega)$ the
instability may occur at $\omega=-1$   \footnote{this is the point we do
theoretically mark as the phantom barrier, beyond which universe may turn
towards different singularities like Big Rip(Caldwell et. al 2003; Nojiri et.
al. 2005) etc.} or  at $\omega=0$ for which the corresponding $z$ is $-0.019$ \&
$52.517$ respectively. However, there are many objects which are at a redshift
of $z=52.517$. But we are yet to speculate any kind of impacts of thermal phase
transition of universe upon them. So based on current knowledge of explaining
observational data we can rule out such a transition at $z=52.517$. Besides, the
model is valid for $z\leq 4$ (Efstathiou, G. 1999). So we can not comment on the
second possibility staying within the model.

Note there is observation constraint on $\omega_{0}$ as follows: $\omega_{0} >
-1$. So we cant have $\omega < -1$ unless we make z slightly negative i.e we are
heading towards instability. Also note $\omega_{1}$ is assumed to be positive to
reconcile with the fact that $\omega$ should decrease with time.

The expression for speed of sound (\ref{12}) within this model also shows
instability since with  $\omega$ going below $-1$, the square of the speed
becomes less than $0$. {\it This is generally termed as adiabatic instability.}
(Bean, R et. al. 2008a, 2008b). Bean, R. et al. (2008a, 2008b) concludes the
models with nontrivial effective coupling between dark matter and
dark energy can lead to exponential growth of small adiabatic
instability which is also characterized by negative sound speed
squared. As a result even if universe starts with a uniform fluid the
instability will bring upon an exponential growth of small density
perturbation. In analogy we also can speculate that it might be possible that at
our indicated points small density perturbation will occur opposing fundamental
phenomenon like the propagation of sound to take place.\\\\
{\bf ASSS Model}\\
Within ASSS model, the expression for $c_{V}$ depicts that the onset of
instability happens when $\omega$ goes below $-2$, which in turn implies the
following timescale of onset of instability:
Taking allowed values of parameter (Alam, U. et. al. 2004a, 2004b) i.e  taking
$A_{0}=0, A_{1}=-1.169, A_{2}=1.67$ we get that the instability happens to set
on at $z=-0.02$. The explanation of this scenario will follow the previous
discussion.

\section{On the Validity of Thermodynamical Laws}\label{section5}
Study of thermodynamical laws with the universe as a thermodynamical system has
been done in any literature(Setare, M.R. 2006, 2007a; Setare, M.R., Shafei, S.
2006; Setare, M.R., Vagenas, E. C. 2008; Mazumder, N., Chakraborty, S. 2009,
2010; Bhattacharya, S. Debnath, U. 2011). We will just recall the result
relevant for our models.
\subsection{Validity on Apparent Horizon}
The Friedman metric of a isotropic spatially homogeneous universe is given by:
\begin{equation}\label{59}
ds^{2}=-dt^{2}+\frac{a^{2}}{1-kr^{2}}dr^{2}+r^{\prime 2}
\left(d\theta^{2}+\sin^{2}{\theta}d\phi^{2}\right)
\end{equation}
where $r^{\prime}=ar$.

From this metric we can easily calculate the radius of the apparent horizon
($r^{\prime}_{AH}$) which comes out to be :
\begin{equation}\label{60}
r^{\prime}_{AH}=ar_{AH}=\frac{1}{\sqrt{H^{2}+\frac{k}{a^{2}}}}
\end{equation}
Now, in spatially flat universe, $k=0$ and thereby we have
$r^{\prime}_{AH}=\frac{1}{H}$.

Following Hawking's idea, the temperature associated with the apparent horizon
is:
\begin{equation}\label{61}
T=\frac{\kappa}{2\pi} 
\end{equation}
where $\kappa$ is surface gravity of the apparent horizon and given by;
$\kappa=\frac{1}{r^{\prime}_{AH}}$.

Hence, the temperature associated with apparent horizon is $\frac{1}{2\pi
r^{\prime}_{AH}}$.

With k=0, using equation (\ref{5}) we obtain
\begin{equation}\label{62}
H^{2}=\frac{8\pi}{3}\rho
\end{equation}
To determine the validity of first law thermodynamics we first calculate the
energy crossing over this apparent horizon in an infinitesimal dt time which
comes out to be (Bousso, R. 2006) 
\begin{equation}\label{63}
-dE_{AH}=\frac{4\pi (P+\rho)}{H^{2}}dt=-\frac{3}{2}\frac{(1+\omega)}{H(1+z)}dz
\end{equation}
On the apparent horizon,the entropy is:
\begin{equation}\label{64}
S_{AH}=\frac{A}{4}=\pi r^{\prime 2}_{AH}
\end{equation}
Hence the following holds :
\begin{equation}\label{65}
T_{AH}dS_{AH}=dr^{\prime}_{AH}
\end{equation}
\begin{equation}\label{66}
dr^{\prime}_{AH}=-\frac{1}{H^{2}}dH=-\frac{3}{2}\frac{(1+\omega)}{H(1+z)}dz
\end{equation}
\begin{equation}\label{67}
\Rightarrow -dE_{AH}=T_{AH}dS_{AH}
\end{equation}
Hence the first law of Thermodynamics holds on apparent horizon independent of
how we are gonna model $\omega(z)$.

Now, Using Gibbs equation we get (Izquirdo, G. and  Pavon D. 2006)
\begin{equation}\label{68}
TdS_{I}=dE_{I}+PdV
\end{equation}
where $S_{I}$, $E_{I}$ are entropy and enrgy density respectively inside the
apparent horizon and $E_{I}$ is given by
\begin{equation}\label{69}
E_{I}=\frac{4}{3}\pi r^{\prime 3}_{AH}\rho =\frac{1}{2}r^{\prime}_{AH}
\end{equation}
and the volume bounded by the apparent horizon is
\begin{equation}\label{70}
V=\frac{4}{3}\pi r^{\prime 3}_{AH}
\end{equation}
Therefore using equation (\ref{68}),(\ref{69}) and (\ref{70}), we get
\begin{equation}\label{71}
dS_{I}=\pi r^{\prime}_{AH}(1+3\omega)dr^{\prime}_{AH}
\end{equation}
Rewriting the last equation and using the expression for $dS_{AH}$,we get
\begin{eqnarray}\label{72}
\frac{d\left(S_{I}+S_{AH}\right)}{dz}=-\frac{9\pi}{2H^{2}}\frac{
\left(1+\omega\right)^{2}}{1+z}\\
\Rightarrow
\frac{d\left(S_{I}+S_{AH}\right)}{da}=\frac{9\pi}{2H^{2}}\left(1+\omega\right)^{
2}\left(1+z\right) > 0
\end{eqnarray}
Hence holds the Generalized Second Law of Thermodynamics if we consider apparent
horizon and the volume bounded by it. So the result holds not only for two types
of fluid we have discussed in the paper, but also for every fluid obeying
$p=\omega(z)\rho$ which is in quite agreement with (Mazumder, N. et. al. 2010).
\subsection{Invalidity of Thermodynamic laws on Event Horizon}
The event horizon $r_{EH}$ is defined to be 
\begin{equation}\label{73}
r_{EH}=\int_{t}^{\infty}\frac{dt}{a}=\frac{1}{1+z}\int_{z}^{-1} \frac{-1}{H} dz
\end{equation}
Hence the event horizon has a hawking's temperature of  $\frac{1}{2\pi r_{EH}}$
which yields
\begin{equation}\label{74}
T_{EH}dS_{EH}=dr_{EH}=\frac{-1}{\left(1+z\right)H}dz
\end{equation}
It follows that:
\begin{equation}\label{75}
dE_{EH}+T_{EH}dS_{EH}=\frac{dz}{\left(1+z\right)H}\left(\frac{3}{2}
\left(1+\omega\right)-1\right) \neq 0 \ unless\ \omega=-\frac{1}{3}
\end{equation}
Note $\omega=-\frac{1}{3}$ is the marginal point of Strong Energy Condition
which is $ 3p + \rho \geq 0$.\\

Employing similar technique we would get, (Xing, L. et. al. 2011) 
\begin{equation}\label{78}
\frac{d\left(S_{I}+S_{EH}\right)}{dz}=2\pi
r_{EH}^{4}H\frac{dH}{dz}+\pi\left[3r_{EH}^{3}H^{2}\left(1+\omega\right)+2r_{EH}
\right]\frac{-1}{\left(1+z\right)H}
\end{equation}
Changing the variable to $a$ we get;
\begin{equation}\label{79}
\frac{d\left(S_{I}+S_{EH}\right)}{da}=2\pi
\left(r_{EH}^{4}H\right)\left(\frac{dH}{da}\right)+\pi\left[3r_{EH}^{3}H^{2}
\left(1+\omega\right)+2r_{EH}\right]\frac{1}{aH}
\end{equation}\\
Evidently $\frac{d\left(S_{I}+S_{EH}\right)}{da}$ is not necessarily positive
for any allowed value of $z$ ( as we have $\frac{dH}{da}$ in the expression)
which, in turn, implies that the second law of thermodynamics breaks down inside
the event horizon in quite agreement with Xing, L. et. al. 2011).
\section{Brief Summary}\label{section6}
So far we have studied the thermodynamics of universe from two different models.
It has been revealed that the first model where
$\omega(z)=\omega_{0}+\omega_{1}ln(1+z)$ does not go through any critical point.
In this model Universe started with a high (fig. 1, 2, 3) energy density,
pressure and temperature which falls as $z$ becomes $2$ from $0$.We imposed
restriction on allowed values of $\omega_{0}$ and $\omega_{1}$ by demanding
$\mathit{v}_{s} \leq 1$. The thermal capacity at constant volume changes sign as
$\omega$ becomes smaller than $-1$, but this transition is smooth without having
critical point. Unlike MCG or GCG here pressure depends both on temperature and
volume in thermal EoS. From equation (\ref{20}) we have shown it is possible to
have $c_{V} > 0$ irrespective of model to ascertain a stable universe. 

Thereafter we have generalized the results for a general EoS parameter. We have
shown that energy density takes the same form as a function of temperature and
volume irrespective of model with a general limiting behavior of thermodynamic
parameter (\ref{57.11}) to satisfy third law of thermodynamics. We have also
interpreted $U_{0}$ as a measure of number of micro states and argued for it
natural plausibility (\ref{57.3}) and (\ref{57.4}). Moreover, it is intriguing
to note that the thermal EoS (\ref{57}) looks like ideal gas EoS in some way.
Also we have determined the asymptotic behavior of $F(V)$ such that third law
of thermodynamics holds. It has been shown that the first and second law of
thermodynamics hold on apparent horizon not only for our model but also for any
model having $p=\omega(z)\rho$. They fail to be valid on event horizon unless
$\omega$ takes a specific value which is in perfect agreement with (Mazumder, N.
et. al.; Xing, L. et. al. 2011).

One important aspect which have been followed in this paper is a try to
speculate the instabilities. Considering, the thermodynamical point at which the
heat capacity changes its sign, as a transition point we have followed that the
point is nothing but the phantom barrier in the case of Log parametrization.
Even there we can find the squared sound speed turns out to be negative. Even
for ASSS such point arises for negative $z$, theoretically indicating to some
future point. We have also explored the possibility of the fact that at our
indicated points small density perturbation will occur such that the fundamental
phenomenon like the propagation of sound does not take place at all.\\\\

{\bf Acknowledgement : }\\\\
SP would like to acknowledge a debt of gratitude to Indian Academy of Science
for financial support and warm hospitality and IISc, Bnagalore for providing
research facilities as the work was done during a time of summer visit. RB thanks
ISRO grant ``ISRO/RES/2/367/10-11" for providing Research Associate Fellowship.
Authors are thankful to Prof Banibrata Mukhopadhyay for fruitful discussions.

\end{document}